\documentclass[aps,prb,twocolumn,showpacs,floatfix]{revtex4}
\usepackage{amsmath,amssymb}
\usepackage{bm}
\usepackage{graphicx}

\begin{document}

\title
{Excitation energy transfer between closely spaced
multichromophoric systems:\\ Effects of band mixing and intraband
relaxation}

\author{C. Didraga}

\author{V.\ A.\ Malyshev}
\thanks{On leave from S.I. Vavilov State Optical Institute,
Birzhevaya Liniya 12, 199034 Saint-Petersburg, Russia.}

\author{J.\ Knoester\footnote[2]
{Corresponding author. Fax: 31-50-3634947. E-mail: j.knoester@rug.nl}}

\affiliation{Institute for Theoretical Physics and  Materials
Science Centre, University of Groningen, Nijenborgh 4, 9747 AG
Groningen, The Netherlands}

\date{\today}

\begin{abstract}

We theoretically analyze the excitation energy transfer between
two closely spaced linear molecular J-aggregates, whose excited
states are Frenkel excitons. The aggregate with the higher (lower)
exciton band edge energy is considered as the donor (acceptor).
The celebrated theory of F\"orster resonance energy transfer
(FRET), which relates the transfer rate to the overlap integral of
optical spectra, fails in this situation. We point out that in
addition to the well-known fact that the point-dipole
approximation breaks down (enabling energy transfer between
optically forbidden states), also the perturbative treatment of
the electronic interactions between donor and acceptor system,
which underlies the F\"orster approach, in general loses its
validity due to overlap of the exciton bands. We therefore propose
a nonperturbative method, in which donor and acceptor bands are
mixed and the energy transfer is described in terms of a
phonon-assisted energy relaxation process between the two new
(renormalized) bands. The validity of the conventional
perturbative approach is investigated by comparing to the
nonperturbative one; in general this validity improves for lower
temperature and larger distances (weaker interactions) between the
aggregates. We also demonstrate that the interference between
intraband relaxation and energy transfer renders the proper
definition of the transfer rate and its evaluation from
experiment a complicated issue, which involves the initial
excitation condition.

\end{abstract}

\pacs{
71.35.Aa;   
73.63.-b;   
78.67.-n    
81.16.Fg    
}

\maketitle

\section{Introduction}
\label{introduction}

The theory of F\"orster resonance energy transfer (FRET) between
two chromophores (molecules, ions) with dipole-allowed optical
transitions,~\cite{Foerster48} and its generalization by
Dexter~\cite{Dexter53} to forbidden optical transitions and
exchange interactions between the chromophores, already have a
history of more than 50 years. This celebrated theory gives an
excellent description of transfer rates for distant chromophores
with rather broad spectral
lines.~\cite{Ermolaev77,Agranovich82,Andrews99} It describes these
rates in terms of the overlap integral of experimentally measured
optical absorption and luminescence spectra, which makes it of
great utility. With minor reformulations, the concept of FRET may
also be applied successfully to the description of nonradiative
transitions in ions and molecules in condensed
phases,\cite{Ermolaev96} as well as to energy transfer in the
presence of a nonstationary bath relaxation.\cite{Jang02} Finally,
it has been shown that the F\"orster theory also explains the
efficient long-range energy transfer in assemblies of closely
packed CdSe quantum dots~\cite{Kagan96,Crooker02,Wargnier04} and
CdSe nanocrystals assembled with molecular wires,\cite{Javier03}
systems of possible use for quantum
computation.\cite{Imamoglu99,Biolatti00}

In spite of its great success, it has been recognized since the
1980's that in certain situations standard FRET theory is not
applicable. In particular, this holds for chromophores with narrow
spectral lines and a small spectral overlap, such as rare-earth
ions embedded in a crystalline or glassy
host.\cite{Holstein81,Malyshev85,Basiev87,Xia02} More recently
another important situation in which FRET theory may break down
has been emphasized, namely energy transfer between two systems
that both contain many interacting chromophores. This problem has
drawn particular attention in the context of excitation energy
transfer from the B800 to the B850 ring of the photosynthetic
antenna system LH2.~\cite{Sumi99,Mukai99,Scholes01,Jordanidis01,%
Scholes03,Jang04,Fleming04} The first complication when dealing
with systems of strongly interacting chromophores is that their
excited states are excitons, consisting of a coherent
superposition of the excited states of many molecules. Due to
their spatial extent, which may easily exceed the separation
between the two systems, the effective interaction between an
exciton state on the donor system and one on the acceptor system
cannot be modelled as the interaction between the transition
dipoles of both states. As a result, excitation energy transfer
may occur from or towards a dipole forbidden (optically dark)
exciton state, implying that a description of the energy transfer
in terms of overlap integrals of optical spectra no longer holds. This
breakdown of the point-dipole approximation was first pointed out
by Sumi and coworkers.\cite{Sumi99,Mukai99} Treating the
electronic coupling between both rings in LH2 as a perturbation,
they derived a transfer rate between the rings that strongly
differed in magnitude from the F\"orster result and which was in
good agreement with experiment.~\cite{Shreve91,Jimenez96}

It should be noted that LH2 is a rather special case of transfer
between two aggregates. The reason is that the molecules in the
B800 ring are weakly coupled to each other. Thus, the B800
excitations are almost monomeric~\cite{vanOijen99} and they occur
in a narrow band just above the upper edge of the B850 band and
far (965 cm$^{-1}$) away from the optically dominant bottom of that
band.~\cite{Mukai99,Wu97} Moreover, the intermolecular
interactions between both rings are weak, in the order of 20
cm$^{-1}$.\cite{Wu97} Given these special circumstances, it is not
surprising that the perturbative treatment of the inter-aggregate
(inter-ring) interactions gives good results.

In many cases of closely separated aggregates, however, a
perturbative treatment of the interaction that causes energy
transfer between them will not be valid. An interesting example is
the case of nanotubular carbocyanine J-aggregates that have
recently been developed by D\"ahne and
coworkers~\cite{vonBerlepsch00a,vonBerlepsch00b,vonBerlepsch03}
and which have been suggested as building blocks for synthetic
light-harvesting systems. These aggregates consist of two walls,
which are only a few nanometers apart. Each wall is responsible
for the formation of an exciton band; even though their optically
dominant lower edges are separated by a few hundred cm$^{-1}$, the
two bands overlap over a large energy range ($\sim$ 2000
cm$^{-1}$).~\cite{Didraga04} Using fluorescence and pump-probe
experiments, fast excitation energy transfer between both walls
has been observed.\cite{Pugzlys04} The strong overlap of
the exciton bands makes it doubtful that a perturbative treatment
of the interwall interactions holds for this example: (dark)
states inside both bands may be close to degenerate, thus falling
outside the perturbative regime. This observation also holds for
the example of energy transfer between two linear pseudo-isocyanine
aggregates studied by Kobayashi and coworkers.~\cite{Fukutake02}
Like Sumi and coworkers, these authors focused on a breakdown of
the dipole approximation; they did not consider the limitations of
a perturbative treatment.

The aim of this paper is to study theoretically the energy
transfer between two molecular J-aggregates carrying Frenkel
excitons. We will be inspired by the example of the double-wall
cyanine tubes, where the bottoms of both exciton bands occur at
different energies, but their central parts overlap. To keep the
problem computationally tractable, we will consider two
interacting linear J-aggregates with different bandwidths, leading
to a crossing of both bands in their center. The chain with the
higher (lower) band bottom is considered the donor (acceptor).
Using this generic model, we will investigate the breakdown of the
perturbative approach by comparing to an exact treatment, in which
both aggregates form one exciton system and the energy transfer is
associated with phonon-assisted relaxation within this system. We
thus find that the crossover between the weak coupling
(perturbative) and strong coupling (nonperturbative) situations is
determined by the separation between both aggregates as well as by
the temperature.
\begin{figure}[th]
\centerline{\scalebox{0.65}{\includegraphics{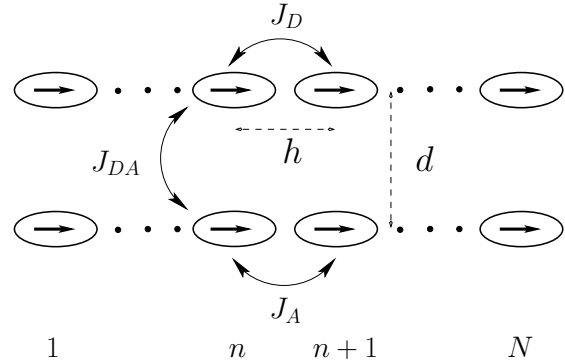}}}
\caption{The two linear J-aggregates (D and A) considered in this
paper. Both chains contain $N$ chromophores, labeled $n$, which
interact with each other through their transition dipoles
(indicated by arrows). The scale of the interactions is set by the
quantities $J_D$, $J_A$, and $J_{DA}$, which give the interactions
between the various pairs of nearest neighbors. The lattice
spacing within each chain is denoted $h$, while the interchain
distance is given by $d$.} \label{cartoon}
\end{figure}

Another important issue which we address is that fast intraband
relaxation (thermalization) between visible and dark exciton
states may obscure the observation of the actual energy transfer
process and thereby plays a crucial role in the proper definition
of the transfer rate as extracted from experiment. As a
consequence, also the initial excitation conditions strongly
affect the possibility to measure the transfer rate. This key role
of intraband relaxation in the process of energy transfer between
excitonic systems seems to have gone unnoticed thus far.

The outline of this paper is as follows. In Sec.~\ref{model} we
first present our model of two homogeneous linear J-aggregates,
interacting with each other as well as with a vibrational bath. We
then introduce the general issue of perturbative versus
nonperturbative treatment of interchain interactions by
considering the various dynamic processes that can occur in and
between weakly coupled or strongly mixed exciton bands. The
perturbative approach is worked out in more detail in
Sec.~\ref{perturbative}, where we derive the general expression
for the one-phonon assisted transfer rate between any two exciton
states located on different chains. In Sec.~\ref{exact} we develop
the approach for the case of strong interchain coupling
(nonperturbative case). Results of numerical simulations of the
fluorescence kinetics in both the perturbative and the
nonperturbative approach are presented in Sec.~\ref{discussion}
and are used to study the validity of the former as well as the
best way to extract the transfer rate from experiment. Both the
temperature and distance dependence of the transfer dynamics are
addressed and distinction is made between resonant and
off-resonant initial excitation of the donor. A comparison to
standard FRET theory is made as well. Finally, in
Sec.~\ref{summary} we summarize.

\section{Model and general strategy}
\label{model}

\subsection{Model}

We consider two parallel linear chains each consisting of $N$
equidistant two-level chromophores, with their transition dipoles
aligned to the chains (see Fig.~\ref{cartoon}). The lattice
spacing within each chain is denoted $h$, while the distance
between the chains is $d$. One of the chains will be referred to
as the donor (D), the other as the acceptor (A) (see below). We
will assume that both chains are homogeneous and we will impose
periodic boundary conditions in the chain direction (the formalism
may easily be extended to account for disorder and open boundary
conditions). The chromophores building up the D chain are
different from those of the A chain; in particular, we will assume
that the transition energies and dipoles of the individual
chromophores within the D chain all have the values
$\varepsilon_D$ and $\mu_D$, respectively, while in the A chain
they take the values $\varepsilon_A$ and $\mu_A$. We will account
for the dipole-dipole interactions between all chromophores in
both chains and also include in the model a coupling of the
electronic excitations to a bath of vibrations. The latter
coupling is derived from the first-order change of the
chromophores' transition energies caused by nuclear displacements
in the environment. In the site representation, the resulting
Hamiltonian of system and bath reads
\begin{eqnarray}
\label{H}
    H & = & H_D + H_A + H_{DA}
\nonumber\\
\nonumber\\
    & + & H_\mathrm{bath} + H_{D-\mathrm{bath}} + H_{A-\mathrm{bath}}\ ,
\end{eqnarray}
with
\begin{subequations}
\begin{equation}
\label{HD}
    H_D = \varepsilon_D \sum_{n = 1}^{N} |n,D\rangle \langle n,D|
    + \sum_{n,m = 1}^{N} J^{D}_{nm} |n,D\rangle \langle m,D|  \ ,
\end{equation}
\begin{equation}
\label{HA}
    H_A = \varepsilon_A \sum_{n = 1}^{N} |n,A\rangle \langle n,A|
    + \sum_{n,m = 1}^{N} J^{A}_{nm} |n,A\rangle \langle m,A|  \ ,
\end{equation}
\begin{equation}
\label{HDA}
    H_{DA} = \sum_{n,m = 1}^{N} J^{DA}_{nm} |n,D\rangle \langle m,A|
    + h.c.  \ ,
\end{equation}
\begin{equation}
\label{Hbath}
     H_\mathrm{bath}=\sum_q \omega_q a^{\dagger}_q a_q \ ,
\end{equation}
\begin{equation}
\label{HD-bath}
    H_{D-\mathrm{bath}} = \sum_{n=1}^N \> \sum_q V^{D}_{nq}
    |n,D\rangle \langle n,D|a_q + h.c.  \ ,
\end{equation}
\begin{equation}
\label{HA-bath}
    H_{A-\mathrm{bath}} = \sum_{n=1}^N \> \sum_q V^{A}_{nq}
    |n,A\rangle \langle n,A|a_q + h.c.  \ .
\end{equation}
\end{subequations}

Here, $H_D$ and $H_A$ denote the electronic (exciton) Hamiltonians
for the isolated donor and acceptor chains, respectively, while
$H_{DA}$ is the electronic interaction between both chains,
composed of all dipole-dipole interactions between molecules of
one chain and the other. In these terms, $|n,D\rangle$
($|n,A\rangle$) denotes the state in which the $n$th
($n=1,\ldots,N$) chromophore of the donor (acceptor) chain is
excited and all the other chromophores are in the ground state.
Furthermore, the hopping integrals $J^D_{nm}$, $J^A_{nm}$, and
$J^{DA}_{nm}$ are the various dipole-dipole interaction matrix
elements between chromophores of the same or different chains. For
the geometry considered here, we have $J^D_{nm} = -J_D/|n -m|^3$,
$J^A_{nm} = -J_A/|n -m|^3$, $J_{nn}^D=J_{nn}^A=0$, and
$J^{DA}_{nm} = J_{DA} \big[
1 - 2(n-m)^2 h^2/d^2 \big]/ \big[ 1 + (n-m)^2h^2/d^2 \big]^{5/2}$,
with $J_D \equiv 2\mu_D^2/h^3$, $J_A \equiv 2\mu_A^2/h^3$, and
$J_{DA}\equiv \mu_D \mu_A/d^3$.

$H_\mathrm{bath}$ describes the vibrational modes of the host,
labeled $q$ and with the energy spectrum $\omega_q$ (we set $\hbar
= 1$). The operator $a$ annihilates a vibrational quantum in mode
$q$. Finally, $H_{D-\mathrm{bath}}$ and $H_{A-\mathrm{bath}}$
represent the operators of the exciton-bath coupling of the donor
and acceptor, respectively, where the quantities $V_{nq}^D$ and
$V_{nq}^A$ indicate their strength. We do not provide explicit
expressions for these quantities; rather we will consider them on a
phenomenological basis. Specifically, realizing that in most
experimental studies of aggregates the host is strongly
disordered, we will treat these strengths as stochastic variables
with correlation properties:
\begin{subequations}
\label{<correlationsVDVA>}
\begin{equation}
\label{<VD>}
    \left\langle V^D_{nq} \right\rangle
    = \left\langle V^A_{nq} \right\rangle
    = \left\langle V^D_{nq}V^{A*}_{mq} \right\rangle = 0 \ ,
\end{equation}
\begin{equation}
\label{<VDVD>}
    \left\langle V^D_{nq}V^{D*}_{mq} \right\rangle
    = \delta_{nm} {\left|V^D_q\right|}^2  \ ,
\end{equation}
\begin{equation}
\label{<VAVA>}
    \left\langle V^A_{nq} V^{A*}_{mq} \right\rangle
    = \delta_{nm} {\left|V^A_{q}\right|}^2  \ .
\end{equation}
\end{subequations}
These relations imply that the surroundings of different
chromophores are not correlated.

\subsection{Strategy} \label{strategy}

We now turn to a general discussion of perturbative versus
nonperturbative approach to describe the excitation energy transfer
between both chains. Throughout this paper, we will assume that
the exciton-phonon coupling is weak compared to the intrachain
dipole-dipole interactions, in the sense that the coherence length
of the excitons within each chain is not limited by the coupling
to the bath. In that case, the Bloch eigenstates of $H_{D}$ and
$H_{A}$ are a good starting point for our considerations and the
main role of the bath is to make up for energy differences in
possible intraband energy relaxation. The Bloch states read
($X=D,A$)
\begin{equation}
\label{eigenstates}
    |k,X \rangle=\sum_{n=1}^{N} \varphi_{k n}^X|n,X\rangle =
    \frac{1}{\sqrt{N}}\sum_{n=1}^{N}
    \exp \left[\frac{2 \pi i k n}{N} \right] |n,X\rangle\ ,
\end{equation}
with energy
\begin{equation}
E_k^X=\varepsilon_X -2J_X \sum_{n=1}^{N/2}
\frac{1}{n^3}\cos\left[\frac{2 \pi i k n}{N} \right].
\label{dispersion}
\end{equation}
Here, $k$ is the wavenumber of the state, which can take the values
$0,1,\ldots,N-1$.

\begin{figure*}[ht]
\centerline{\scalebox{0.95}{\includegraphics{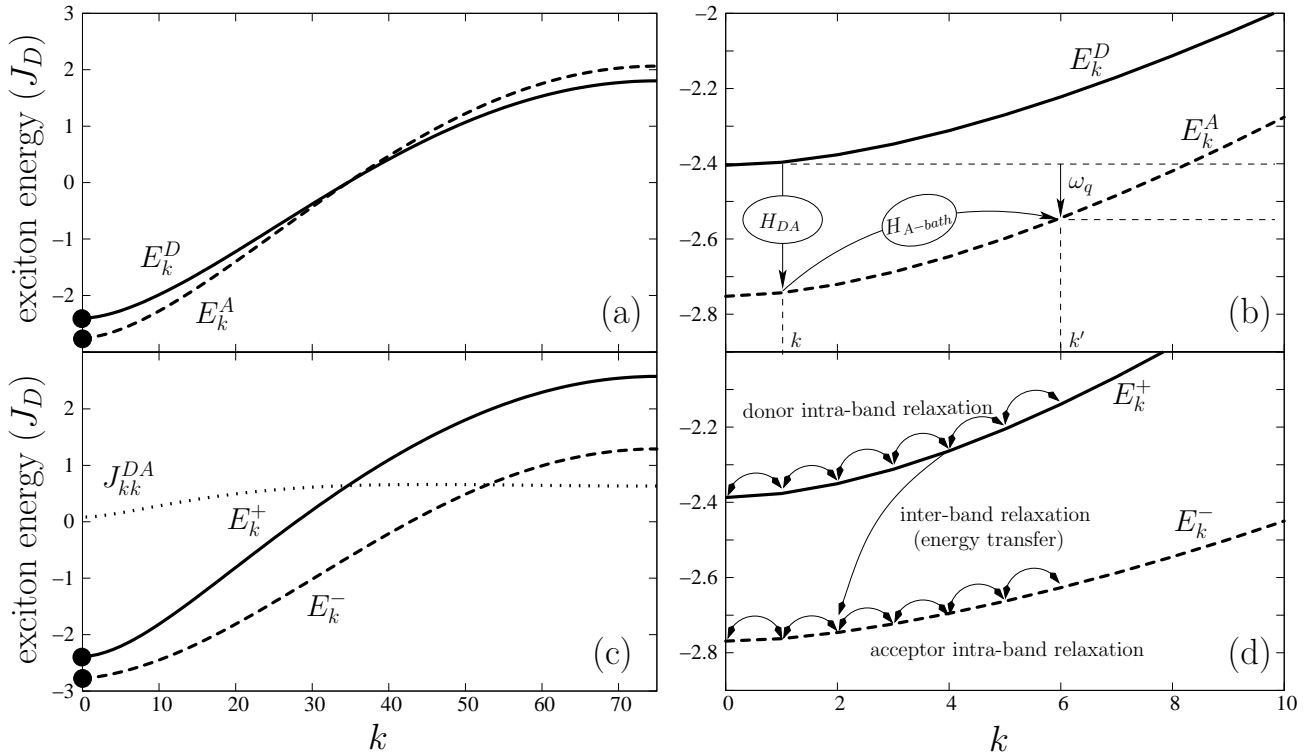}}}
\caption{(a) Exciton bands for donor (solid) and acceptor (dashed)
chains of $N=150$ molecules, in the absence of interchain coupling
for the case $J_A=1.14J_D$ and setting
$\varepsilon_A=\varepsilon_D=0$. Note that we have only plotted
half of the Brillouin zone, as it is symmetric around $k=N/2$. The
dots at $k=0$ indicate the positions of the absorption peaks of
donor and acceptor band, respectively. (b) Expansion of the
small-$k$ part of panel (a), with schematic indication of the
two-step perturbative view of the interchain energy transfer
described in the text. (c) Exciton bands for the same system as in
panel (a), but now after accounting for the mixing between the
eigenstates of both chains due to the interchain interaction with
strength $J_{DA}=0.535J_D$ (corresponding to an interchain
separation $d=h$.) The $k$-dependence of the interchain coupling
$J^{DA}_{kk}$ is depicted as well. (d) Expansion of the small-$k$
part of panel (c), with schematic indication of the various
phonon-assisted intraband and interband relaxation processes that
together are responsible for the exciton dynamics within the
nonperturbative picture.} \label{fig:bands}
\end{figure*}

Figure~\ref{fig:bands}(a) shows the donor and the acceptor bands,
$E_k^D$ and $E_k^A$, respectively, for chains of $N=150$ molecules,
with $\varepsilon_D=\varepsilon_A$ (we take this as zero of
energy) and $J_A=1.14J_D$. Only half of the Brillouin zone is
plotted, as the dispersion relation is symmetric around $k=N/2$.
We see two exciton bands that cross at their center. The bottom of
each band occurs at $k=0$, which is the superradiant state that
contains all oscillator strength to the ground state; all other
exciton states are optically forbidden. Thus, as long as the
interactions between the chains are weak, the fluorescence
spectrum consists if two J-bands, whose positions are indicated by
the dots on the $k=0$ axis. By definition, we choose our labeling
such that $E_{k=0}^D > E_{k=0}^A$. The energy separation between
both peaks will be denoted $\Delta$ and is (for $N \gg 1$) given by
$\Delta=\varepsilon_D- \varepsilon_A +2\zeta(3)(J_A-J_D)$, with
$\zeta(x)$ the Riemann zeta function.

The two chains are coupled electronically by the interaction
$H_{DA}$ (Eq.~(\ref{HDA})), which on the basis of Bloch states is
diagonal, i.e., $H_{DA}=\sum_k J^{DA}_{kk} |k,D\rangle \langle
k,A|$ +h.c., with
\begin{equation}
\label{JDAkk}
    J^{DA}_{kk} = J_{DA}+2J_{DA} \sum_{n = 1}^{N/2} \,
    \frac{1 - 2 (nh/d)^2 }{\left[ 1
    + (nh/d)^2 \right]^{5/2}} \, \cos\left(\frac{2\pi
    kn}{N}\right)\ .
\end{equation}
In general, this expression must be evaluated numerically (see
Fig.~\ref{fig:bands}(c) for an example). Only in the limiting (and
rather unphysical) case of $d \ll h$ it is easily seen that
$J^{DA}_{kk} = J_{DA}=\mu_D \mu_A/d^3$, independent of $k$.

As long as the interactions $J^{DA}_{kk}$ are weak (large
separation $d$), it seems reasonable to apply the usual
perturbative approach to them, using the bands of the isolated
donor and acceptor chains as starting point. We will follow this
route in Sec.~\ref{perturbative}, where we also will treat the
exciton-phonon interactions perturbatively. In this approach, the
rate of energy transfer between an arbitrary donor state and
acceptor state is calculated using second order perturbation
theory, involving two steps: (i) The excitation is transferred
from the donor to the acceptor. (ii) A phonon-induced scattering
occurs within the acceptor band (scattering within the donor
followed by transfer to the acceptor is possible as well). The
overall process should conserve the total energy. Both steps are
schematically indicated in Fig.~\ref{fig:bands}(b), in which we
zoomed in on the small-$k$ part of the bands given in
Fig.~\ref{fig:bands}(a). As in our example the interchain
interaction is diagonal in $k$, the transfer step is a vertical
transition between both bands. We notice that aside from energy
transfer between both chains, the interactions with the phonon
bath also give rise to relaxation of the exciton states within
both bands; details of this process are studied in
Ref.~\onlinecite{Bednarz02}.

If we focus on the optically dominant bottom states of both bands,
it seems that the perturbative treatment of the  interchain
interactions is valid as long as $|J^{DA}_{k=0,k=0}| \ll \Delta$.
Generally speaking, this criterion is not sufficient, however.
Depending on temperature and initial excitation condition, the
transfer process may involve the optically forbidden states higher
in the exciton bands. As near the band center the donor and
acceptor states get arbitrarily close in energy, a perturbative
approach necessarily fails there. For this situation, one has to
resort to a nonperturbative treatment, in which both bands are
mixed. As long as $|J^{DA}_{k=0,k=0}| \ll \Delta$, the amount of
mixing will be weak for the superradiant band bottoms, but strong
near the center.

For our case of translational symmetry, the mixed eigenstates of
$H_D+H_A+H_{DA}$ read (see, e.g., Ref.~\onlinecite{LandauQM})
\begin{subequations}
\label{eigenproblem}
\begin{eqnarray}
\label{pm}
    |k,\pm \rangle = \frac{1}{\sqrt{2}} \left[
    \left(1 \pm \frac{1}{\sqrt{1 + \eta_k}} \right)^{1/2}|k,D \rangle
\right.
\nonumber\\
\nonumber\\
\left.
    \pm \left(1 \mp \frac{1}{\sqrt{1 + \eta_k}} \right)^{1/2}|k,A \rangle
    \right] \ ,
\end{eqnarray}
with energies
\begin{equation}
\label{Epm}
    E^{\pm}_{k} = \frac{1}{2}\left( E^D_k + E^A_k \right)
    \pm \frac{1}{2} \left( E^D_k - E^A_k \right)\sqrt{1+\eta_k} \ .
\end{equation}
\end{subequations}
Here, $\pm$ label the new (decoupled) exciton bands and $\eta_k =
4|J^{DA}_{kk}|^2/(E^D_k - E^A_k)^2$ is the quantity that
characterizes the amount of mixing at wavenumber $k$. The new
bands are plotted in Fig.~\ref{fig:bands}(c) for the same
parameters as in Fig.~\ref{fig:bands}(a), accounting for an
interchain interaction of strength $J_{DA}=0.535J_D$ (which
corresponds to a small interchain separation of $d=h$). The
coupling $J^{DA}_{kk}$ is depicted as well. As we see, at the band
bottoms, the coupling is very small compared to $\Delta$ and the
states are hardly mixed ($\eta_{k=0}=0.2$). Hence, we may still
refer to the fluorescence coming from the bottom of the upper
(lower) band as the donor (acceptor) fluorescence. At the center
of the bands, the mixing is very strong and a band anticrossing occurs.
Also after mixing, the only states with oscillator strength occur
at the band bottoms.

After accounting for the band mixing, the only remaining dynamics
that can take place is phonon-assisted scattering of the new
exciton states, leading to intraband and interband relaxation of
the excitation energy (see processes indicated in
Fig.~\ref{fig:bands}(d)). In Sec.~\ref{exact}, this approach will
be further specified, using the Fermi golden rule to account for
the exciton-phonon interaction.

To end this section, we stress that the special conditions imposed
in our model, such as chains of equal length and periodicity,
periodic boundary conditions, and the absence of disorder in the electronic
part of the Hamiltonian, may be
relaxed without affecting the above formalism and the distinction
between the perturbative and nonperturbative approach. In fact, in
all expressions presented in Secs.~\ref{perturbative} and
\ref{exact}, we will use a general notation for the exciton wave
functions, $\varphi_{k n}^D$, $\varphi_{k n}^A$, and $\phi_{\nu
n}$, so that these expressions keep their validity under
generalized conditions, as long as the proper exciton
eigenfunctions of the generalized $H_D$, $H_A$, and $H_D + H_A +
H_{DA}$ are used as input. In that case, $k$ and $\nu$ refer to
the appropriate quantum numbers, which do not necessarily have the
meaning of quasi-momenta.

\section{Perturbative approach}
\label{perturbative}

In this section, we follow the perturbative approach outlined in
Sec.~\ref{strategy}. In this approach, the Hamiltonian of the
unperturbed system reads $H_0 = H_D + H_A + H_\mathrm{bath}$, while
$H^{\prime} = H_{DA} + H_{D-\mathrm{bath}} + H_{A-\mathrm{bath}}$
represents the perturbation that induces transitions between the
eigenstates of $H_0$. On the basis of exciton eigenstates of the
noninteracting chains (Eq.~\ref{eigenstates}), the interchain
coupling and the exciton-bath interactions take the form
\begin{subequations}
\label{H'-excitonic}
\begin{equation}
\label{DA-excitonic}
    H_{DA} = \sum_{k,k^{\prime} = 1}^N J^{DA}_{kk^{\prime}}
    |k,D \rangle\langle k^{\prime},A | + h.c. \ ,
\end{equation}
with $J^{DA}_{k k^{\prime}} = \sum_{n,m} J^{DA}_{nm}
\varphi^{D}_{k n} \varphi^{*A}_{k^{\prime} m}$, and
\begin{equation}
\label{D-bath-excitonic}
    H_{X-\mathrm{bath}} = \sum_{k,k^{\prime} = 1}^N \>
    \sum_q  V^X_{k k^{\prime} q}
    |k,X \rangle \langle k^{\prime},X| a_q + h.c.  \ ,
\end{equation}
with $V^X_{k k^{\prime} q} = \sum_n \> V^X_{nq} \varphi^{*X}_{k n}
\varphi^X_{k^{\prime} n}$ and $X = D,A$.
\end{subequations}

We aim to consider the energy transfer between any donor state
$|k,D \rangle$ and any acceptor state $|k^{\prime},A \rangle$.
Because the transfer operator $H^{\prime}$ does not produce such
transitions within first-order perturbation theory, we have to
resort to the second-order term. The corresponding expression for
the transfer rate reads
\begin{eqnarray}
\label{WkAkD-general}
    W_{k^{\prime}A,\> kD}
    & = & 2\pi \sum_f \sum_i \rho(E_i) \left \langle \left|
    \sum_s \frac{
    \langle f| H^{\prime} |s \rangle
    \langle s|H^{\prime} |i \rangle}
    {E_i - E_s}
    \right|^2 \right\rangle
\nonumber\\
\nonumber\\
    & \times & \delta(E_f - E_i) \ .
\end{eqnarray}
Here, $|i \rangle = |k,D\rangle  |\{n_q\}_i \rangle$ and $|f
\rangle = | k^{\prime},A \rangle |\{n_q\}_f \rangle$ are the
initial and final states, respectively, where $\{n_q\}$ denotes
the set of occupation numbers of the vibrational modes. $E_i =
E_k^D + \Omega_i$ and $E_f = E_{k^{\prime}}^A + \Omega_f$ are the
corresponding energies, with $\Omega_i$ and $\Omega_f$ denoting
the energies of the bath in the initial and final states,
respectively. Furthermore, $s$ labels the intermediate states $|s
\rangle$, with energies $E_s$ and the quantity $\rho(E_i)$ is the
equilibrium density matrix of the bath's initial state. Finally,
the angular brackets indicate that we average over the stochastic
realizations of the surroundings of donor and acceptor
chromophores.

Evaluating the expressions in Eq.~(\ref{WkAkD-general}) and
accounting for the stochastic properties
Eqs.~(\ref{<correlationsVDVA>}) of the exciton-bath couplings, we
obtain
\begin{eqnarray}
\label{WkAkD-explicit}
    W_{k^{\prime}A,\> kD}& = & {\cal D}(|E^D_k - E^A_{k^{\prime}}|)
    \Bigg[ \sum_{k^{\prime\prime}}
    \frac{|J^{DA}_{k^{\prime}k^{\prime\prime}}|^2}
    {\left(E^A_{k^{\prime}} - E^D_{k^{\prime\prime}}\right)^2} \>
    {\mathcal O}^D_{k^{\prime\prime}k}
\nonumber\\
    & + & \sum_{k^{\prime\prime}}{\mathcal O}^A_{k^{\prime}k^{\prime\prime}}
     \frac{|J^{DA}_{k^{\prime\prime}k}|^2}
    {\left(E^D_k - E^A_{k^{\prime\prime}}\right)^2}\Bigg]
\nonumber\\
\nonumber\\
    & \times &  \begin{cases}
    1+ \bar{n}(E^D_k - E^A_{k^{\prime}})\ , &  E^D_k > E^A_{k^{\prime}}  \,,
    \\
    \\
    \bar{n}(E^A_{k^{\prime}} - E^D_k) \ , &  E^D_k < E^A_{k^{\prime}} \ ,
\end{cases}
\end{eqnarray}
where ${\cal D}(\omega)$ is the vibration spectral density of the
bath, which we assume to be identical for the donor and acceptor.
It is given by
\begin{equation}
\label{D}
    {\cal D}(\omega) = 2\pi \sum_q \left|V^X_q \right|^2
    \delta(\omega - \omega_q) \ ,
\end{equation}
where $X = D,A$. The quantity ${\mathcal O}^X_{kk^{\prime}}$
denotes the probability overlap of the donor states $|k,D \rangle$
and $|k^{\prime},D \rangle$ for $X=D$ and of the acceptor states
$|k,A \rangle$ and $|k^{\prime},A \rangle$ for $X=A$:
\begin{equation}
\label{overlap}
    {\mathcal O}^X_{kk^{\prime}} = \sum_{n=1}^N
    |\varphi^X_{k n}|^2 |\varphi^X_{k^{\prime} n}|^2\,.
\end{equation}
Finally, $\bar{n}(\omega_q) = [\exp(\omega_q/T) - 1]^{-1}$ is the
mean occupation number of the vibrational mode $q$ (the Boltzmann
constant $k_B = 1$).

The transfer rate Eq.~(\ref{WkAkD-explicit}) reflects the two-step
processes introduced in Sec.~\ref{strategy}. The first term
corresponds to the process in which the initial exciton of
wavenumber $k$ on the donor chain is scattered into the exciton
state $k^{\prime\prime}$ on the donor under the creation or
annihilation of a phonon $q$, followed by the transfer of the
exciton $k^{\prime\prime}$ to an exciton $k^{\prime}$ on the
acceptor chain. Likewise, the second term in
Eq.~(\ref{WkAkD-explicit}) derives from the process in which the
donor's initial exciton $k$ is first transferred to the acceptor
chain $k^{\prime\prime}$, followed by a phonon-induced scattering
to the final acceptor state $k^{\prime}$.

Equation~(\ref{WkAkD-explicit}) clearly indicates that in general
the energy transfer may occur between any pair of donor and
acceptor states, independently of whether those states are
optically allowed or forbidden: neither the probability overlap
${\mathcal O}^X_{kk^{\prime}}$ nor the transfer interactions
$J^{DA}_{kk^{\prime}}$ vanish for general $k$ and $k^{\prime}$.
Equation~(\ref{WkAkD-explicit}) reduces to F\"orster's formula
only if the distance $d$ between the chains is larger that their
lengths $Nh$. Indeed, in that case the interchain interactions are
well-approximated by $J^{DA}_{kk^{\prime}} = (\mu_D \mu_A/d^3)
\sum_{n=1}^N \varphi^{*D}_{k n} \sum_{m=1}^N \varphi^A_{k^{\prime}
m}$, where the sums correspond to the dimensionless transition
dipole moments of the states $|k,D \rangle$ and $|k^{\prime},A
\rangle$. For our ordered chains, these dipole moments are giant
for $k = k^{\prime} = 0$, while they vanish for all other states.
Thus, the term $J^{DA}_{k=0,k^{\prime}=0}$ is dominant. The energy
transfer between the chains may then be viewed as F\"orster-type
due to the overlap of the donor's side-band fluorescence and the
zero-phonon acceptor absorption and {\it vice versa}.

The importance of the forbidden exciton states in the energy
transfer between excitonic systems was mentioned for the first
time by Sumi and co-workers~\cite{Sumi99,Mukai99} and later on
received attention from several authors.\cite{Scholes01,Jordanidis01,%
Fukutake02,Scholes03,Jang04,Fleming04} We note that in
Refs.~\onlinecite{Sumi99} and~\onlinecite{Mukai99} the problem was
treated in a more general way than we did above: only the transfer
interactions $J^{DA}_{nm}$ were considered as perturbations, while
the exciton-bath couplings were accounted for by a self-consistent
second-order evaluation of the self-energy. We do not use such a
self-consistent treatment of the exciton-bath interaction in this
paper, as it turns out impossible to combine it with the
nonperturbative treatment of the transfer interactions which we
will investigate in the next section.

The expression for $W_{k^{\prime}A,kD}$ obtained above allows one
to calculate the energy transfer rate between any pair of donor and
acceptor states. From the occurrence of the energy denominators in
Eq.~(\ref{WkAkD-explicit}), however, it is clear that the
perturbative approach of the interchain interactions is restricted
to situations where the donor and acceptor bands do not cross. A
band crossing or (for disordered systems) a band overlap will lead
to small denominators, breaking the result of perturbation theory
and giving rise to rates that are large compared to the energy
spacing of the states involved. This motivates the nonperturbative
approach presented in the next section.

\section{Nonperturbative approach}
\label{exact}

In this section, we work out in more detail the nonperturbative
approach outlined in Sec.~\ref{strategy}, in which the interchain
interactions are accounted for exactly, through mixing the exciton
bands of the isolated chains. We will label the $2N$ mixed
eigenstates of $H_{ex}=H_D + H_A + H_{DA}$ with a greek index
$\nu=1,\ldots,2N$; they take the form
\begin{equation}
\label{Phimu} |\nu\rangle =\sum_{n=1}^{2N} \phi_{\nu n} |n\rangle \, ,
\end{equation}
where $|n\rangle=|n,D\rangle$ for $n=1,\ldots,N$ and
$|n\rangle=|n-N,A\rangle$ for $n=N+1,\ldots,2N$. The corresponding
energy is denoted $E_{\nu}$. For the special highly symmetric model
considered in this paper, the explicit forms of the eigenstates and
energies have been given in Sec.~\ref{strategy} already, where the
$\nu$ index should be identified with the $k\pm$ labeling. In the
current section, the notation is kept general.

In the new representation, the exciton-bath coupling, which is the
perturbation that causes the exciton dynamics, reads
\begin{equation}
\label{H'}
    H_{D-\mathrm{bath}}+H_{A-\mathrm{bath}} = \sum_{\mu,\nu=1}^{2N}
    \sum_{q} V_{\mu \nu q} |\mu\rangle\langle\nu| a_q + h.c. \, ,
\end{equation}
with $V_{\mu \nu q} = \sum_{n=1}^{2N} V_{nq} \phi^*_{\mu n} \phi_{\nu n}$,
where, as before, the coupling strength $V_{nq}$ is considered a
stochastic function of the site index $n$ with properties similar
to those given by Eqs.~(\ref{<correlationsVDVA>}): $\left\langle
V_{nq} \right\rangle = 0$ and $\left\langle V_{mq}V^*_{nq}
\right\rangle = \delta_{mn} {\left|V_q\right|}^2$.

As discussed in Sec.~\ref{strategy}, the dynamics of the excitons
in the new representation is caused by their scattering on
phonons. In order to describe this process, we will use a Pauli
Master Equation for the populations $P_{\nu}(t)$ of the exciton
states:
\begin{equation}
\label{Pnu}
    \dot P_\nu = -\gamma_\nu P_\nu + \sum_{\mu}
    \left( W_{\nu\mu}P_{\mu} - W_{\mu\nu}P_\nu \right) \ .
\end{equation}
Here, $\gamma_\nu = \gamma_0 F_\nu$ is the radiative decay rate of
the state $|\nu \rangle$, where $\gamma_0$ is this rate for a
single chromophore (for simplicity taken identical for donor and
acceptor) and $F_\nu = \big( \sum_{n=1}^{2N} \phi_{\nu n} \big)^2$
denotes the dimensionless oscillator strength of the exciton.
Furthermore, the scattering rates $W_{\mu\nu}$ are obtained using
Fermi's golden rule, taking into account the stochastic properties
of the exciton-bath coupling strength $V_{qn}$. They are given
by~\cite{Bednarz02,Heijs05}
\begin{eqnarray}
\label{Wmunu}
    W_{\mu\nu} &=& \mathcal{D}(|E_\mu-E_\nu|){\mathcal O}_{\mu\nu}
\nonumber\\
\nonumber\\
    && \times \begin{cases}
    1+\bar{n}(E_\nu-E_{\mu}) \ , &  E_\nu > E_{\mu}  \,,
    \\
    \\
    \bar{n}(E_{\mu}-E_\nu) \ , &  E_{\mu} >  E_\nu \,,
\end{cases}
\end{eqnarray}
where ${\cal D}(\omega)$ is the spectral density of the bath, given
by Eq.~(\ref{D}) with $V^X_q$ replaced by $V_q$. Furthermore,
${\mathcal O}_{\mu\nu}$ is the probability overlap of states $\mu$
and $\nu$,
\begin{equation}
\label{overlapD+A}
    {\mathcal O}_{\mu\nu}=\sum_{n=1}^{2N} |\phi_{\mu n}|^2
    |\phi_{\nu n}|^2 \, .
\end{equation}
 The explicit expression
for ${\mathcal O}_{\mu\nu}$ in the case of homogeneous chains of
equal length is given in the Appendix.

To end this section, we note that in general
the best way to probe the excitation energy transfer is to follow
the fluorescence kinetics of donor and acceptor. Even if we allow
for inhomogeneity, two coupled aggregates with a clear separation
between their individual J-bands will lead to a mixed system that
still has two J-bands, possibly renormalized in magnitude and
shifted somewhat (see Fig.~\ref{fig:bands} for the special case of
ordered chains). In terms of the energies of the mixed eigenstates
and the solution to the Pauli master equation, the time-dependent
fluorescence spectrum reads
\begin{equation}
\label{fluorescence}
    I(E,t)=\left \langle \sum_{\nu}\gamma_\nu P_\nu (t) \delta (E - E_\nu)
    \right \rangle \ ,
\end{equation}
where the brackets denote the average over disorder, if present.
One may then define the total donor (acceptor) fluorescence as the
integral over the highest (lowest) peak in the spectrum. These two
quantities will be analyzed in Sec.~\ref{discussion} as a function
of time following some initial excitation of the donor chain.

\section{Results and discussion}
\label{discussion}

We now apply the formalism developed in the previous sections to
study the energy transfer rates and fluorescence kinetics for the
model introduced in Sec.~\ref{model}, namely two parallel
homogeneous chains of $N$ chromophores, with periodic boundary
conditions. In all examples, we will choose $N=150$,
$\varepsilon_A=\varepsilon_D$ and $J_A=1.14J_D$, which implies that
$\mu_A=1.07\mu_D$; the value of $J_{DA}$ depends on the interchain
distance $d$, which we will vary. Given the above parameters, we
have $J_{DA}=(\mu_A/\mu_D)(h/d)^3J_D/2= 0.535(h/d)^3J_D$.

For the spectral density of the bath we choose
\begin{equation}
\label{D-model}
    D(\omega) =
    W_0 \frac{\omega}{\omega_c}
    \exp\left(-\frac{\omega}{\omega_c}\right) \ ,
\end{equation}
which represents a model function with Ohmic (i.e., linear)
behavior~\cite{Weiss} for frequencies up to a cut-off frequency
$\omega_c$. The overall prefactor $W_0$ is a
measure of the exciton scattering amplitude imposed by the
phonons. A spectral density similar to Eq.~(\ref{D-model}), with
$\omega_c$ in the order of 100 cm$^{-1}$, has been used
successfully to fit the optical dynamics in photosynthetic antenna
complexes.~\cite{Kuhn97,May00,Renger01,Brueggemann04} Also,
Eq.~(\ref{D-model}) without a cut-off has been used to explain the
optical dynamics of aggregates of the dye
3,3'-bis(sulfopropyl)-5,5'-dichloro-9-ethylthiacarbo-cyanine
(THIATS) measured between 0 and 100 K.~\cite{Bednarz03} In the
remainder of this paper we will use $\omega_c = 0.2 J_D$, which is
reasonable for J-aggregates, where $J_D$ typically lies in the
range 500-1000 cm$^{-1}$. $W_0$ and $J_D$ will be kept arbitrary,
where $W_0^{-1}$ will serve as unit of time and $J_D$ as unit of
temperature. For example, for $J_D=800$ cm$^{-1}$, the choice
$T/J_D=0.25$ agrees with room temperature.

The above specifies all input necessary to determine the transfer
rate $W_{k^{\prime}A,kD}$ [Eq.~(\ref{WkAkD-explicit})] between
arbitrary states in the perturbative approach, as well as the
relaxation rates $W_{\mu\nu}$ [Eq.~(\ref{Wmunu})] in the
nonperturbative treatment. In the latter case, the transfer from
the donor to the acceptor involves the relaxation rate
$W_{k^{\prime}-,k+}$.

\subsection{Nonperturbative-to-perturbative crossover}
\label{comparison}

In order to assess the validity of the perturbative approach, we
calculate the total transfer rate from donor to acceptor chain
within both approaches, assuming that the initial exciton
populations of the donor manifold are in thermal equilibrium,
i.e., we assume that the intraband relaxation is fast compared to
the energy transfer process. For the nonperturbative approach, this
means that we use as initial condition $P_{k+}(t=0) = Z^{-1}_+
\exp(-E_{k+}/T)$, where $Z_+ = \sum_k \exp(-E_{k+}/T)$, while the
acceptor band is not populated, $P_{k-}(t=0) = 0$. Then, the
effective quantity
\begin{equation}
\label{W+to-}
    W_{-+} = \frac{1}{Z_+} \sum_{k,k^{\prime}}
    \exp\left(-\frac{E_{k+}}{T} \right)
    W_{k^{\prime}-,k+} \ ,
\end{equation}
may be associated with the energy transfer rate from the donor to
the acceptor band at the initial stage of the transfer, when the
back-transfer is negligible. In the perturbative treatment, the
analog of this effective transfer rate is given by
\begin{equation}
\label{W+to-pert}
    W_{AD} = \frac{1}{Z_D} \sum_{k,k^{\prime}}
    \exp\left(-\frac{E^D_k}{T} \right)
    W_{k^{\prime}A,kD} \ ,
\end{equation}
with $Z_D = \sum_k \exp(-E^D_k/T)$. By construction, $W_{-+}
\rightarrow W_{AD}$ in the limit of large interchain distance $d$,
i.e., small interchain interactions ($\eta_k \ll 1$).

\begin{figure}[thb]
\centerline{\scalebox{0.65}{\includegraphics{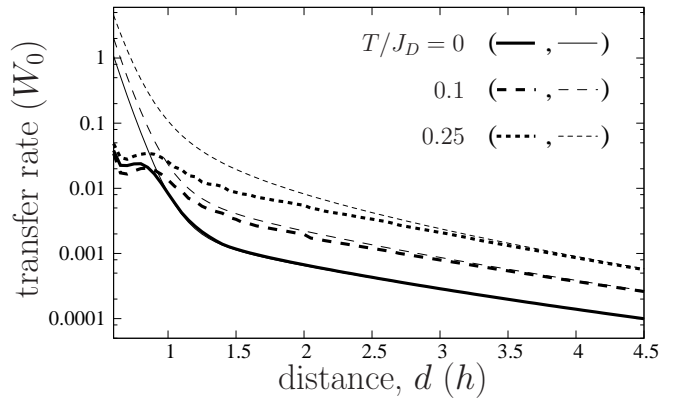}}}
\caption{Donor to acceptor energy transfer rate in the
nonperturbative approach (thick lines) and the perturbative
approach (thin lines), as a function of the interchain distance
$d$ for various temperatures $T$. These rates are given by
$W_{-+}$ [Eq.~(\ref{W+to-})] and $W_{AD}$ [Eq.~(\ref{W+to-pert})],
respectively. All parameters as given at the beginning of
Sec.~\ref{discussion}.} \label{fig:pert-nonpert}
\end{figure}

Figure~\ref{fig:pert-nonpert} shows the results of our calculations
for $W_{-+}$ (thick lines) and $W_{AD}$ (thin lines) as a function
of the interchain distance $d$. From this figure it is clearly
seen that the perturbative approach strongly overestimates the
energy transfer rate for small $d$, as expected. It is also seen
that in the limit of large $d$, both treatments indeed are in
perfect agreement. However, the value of $d$ below which the
perturbative approach fails, increases for growing temperature.
This finds its natural explanation in the band mixing. Let us first
consider the situation at $T=0$. Then only relaxation from the
donor band edge state $|0+ \rangle$ to states in the acceptor band
below the donor band edge contribute to $W_{-+}$. For our choice
of parameters, even at distances as small as $d = h$, these states
in the vicinity of the band edges are only weakly coupled
($\eta_{k \approx 0}$ is rather small, see Sec.~\ref{strategy}).
This explains why the difference between the perturbative and
nonperturbative results quickly vanishes for $d> h$. This
difference only is considerable for $d < h$, because the
interchain interactions then quickly grow ($\sim 1/d^3$),
increasing the band mixing even for the band edge states.

\begin{figure}[thb]
\centerline{\scalebox{0.65}{\includegraphics{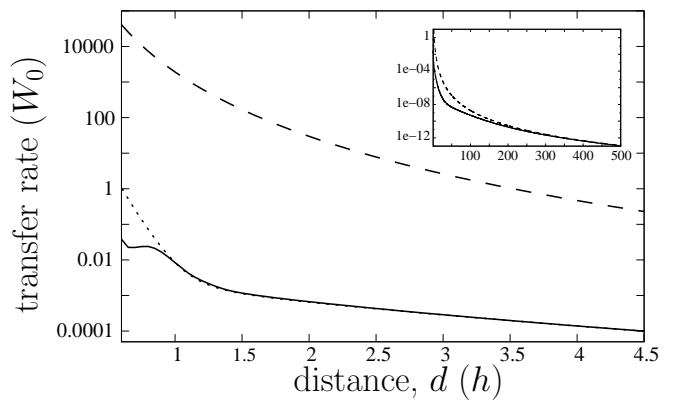}}}
\caption{Energy transfer rate as a function of interchain distance
at $T=0$ according to F\"orster's formula (dashed line, see text
for definition) compared to the exact result [solid,
Eq.~(\ref{W+to-})] and the perturbative one [dashed,
Eq.~(\ref{W+to-pert})]. All parameters as given at the beginning
of Sec.~\ref{discussion}. The inset shows the convergence to
F\"orster's result for very large distances.} \label{fig:Forster}
\end{figure}

Upon increasing the temperature, the initial population will
spread to higher $k$ states in the donor band. For these states,
the band mixing becomes increasingly more important, even for
growing values of the distance $d$. This effect finds its origin in
the fact that the energy separation between the isolated donor and
acceptor bands decreases with growing $k$, while the interchain
interaction $J^{AD}_{kk}$ grows (cf.~Fig.~\ref{fig:bands}). This
explains why the range of $d$ over which the perturbative approach
fails, increases with growing temperature.

It is of interest also to compare the above derived energy
transfer rates with F\"orster's formula, which is obtained from
Eq.~(\ref{W+to-pert}) under the assumption that $J^{DA}_{kk} =
(\mu_D \mu_A/d^3) N \delta_{k0}$ (cf.~Sec.~\ref{perturbative}).
Here, we limit ourselves to the rate at $T=0$, for which
F\"orster's rate reduces to $W_{AD}^F = 1911 W_0 (h/d)^6$. This
result is plotted as a function of $d$ in Fig.~\ref{fig:Forster},
together with $W_{-+}$ and $W_{AD}$ at $T=0$. Clearly, the
F\"orster result gives an enormous overestimation of the exact as
well as the perturbative rate for interchain distances smaller
than roughly the chain length. This is not surprising, because at 
distances small compared to the
chain size, the $J_{kk^{\prime}}^{DA}$ defined below
Eq.~(\ref{DA-excitonic}), does not reduce to the interaction
between the excitons' transition dipoles.

\begin{figure*}[thb]
\centerline{\scalebox{1.0}{\includegraphics{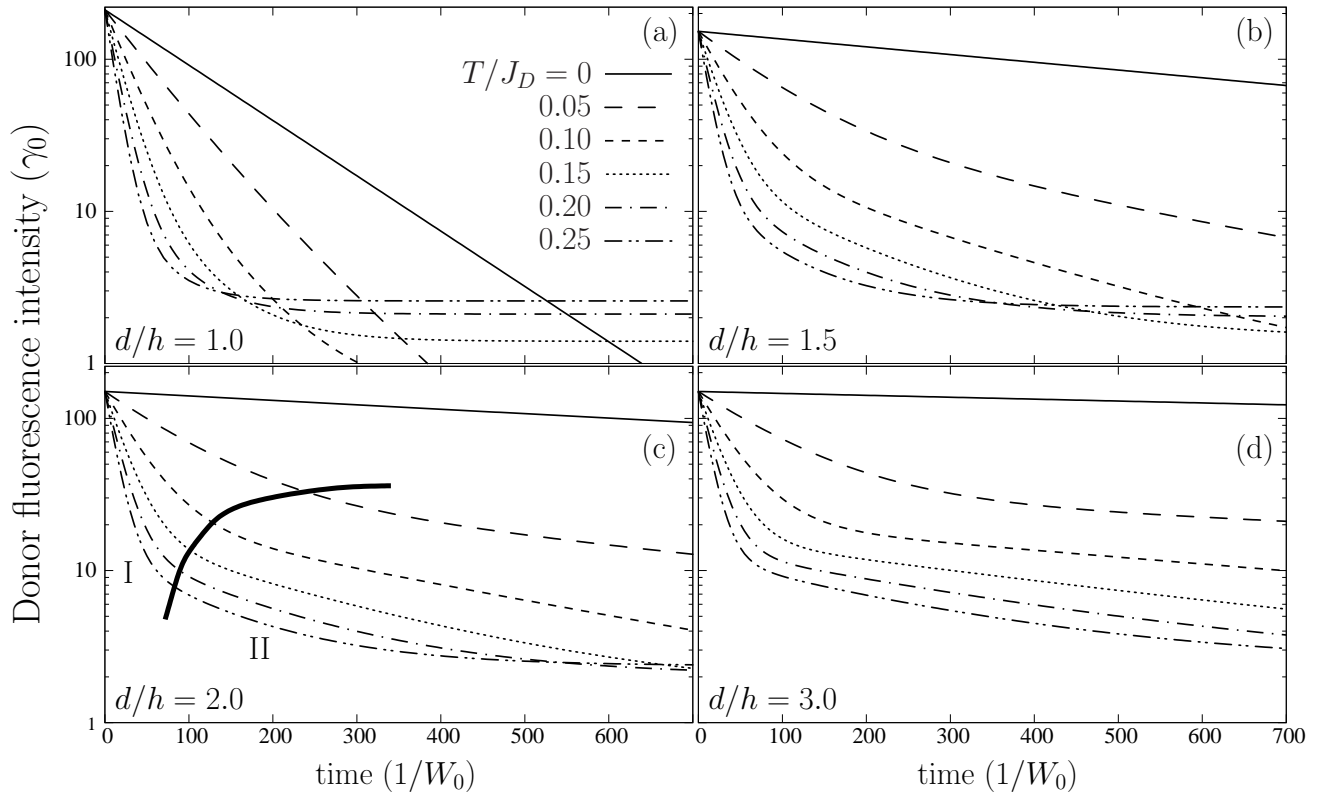}}}
\caption{Kinetics of the donor fluorescence $I_D(t)$ following
resonant excitation $[P_{k+}(0) = \delta_{k0}]$ for different
temperatures $T$ and distances $d$ between the chains, as
indicated in the panels. All other parameters as given at the
beginning of Sec.~\ref{discussion}. The thick line in panel (c)
separates the two kinetic stages (I and II) distinguished in the
text.} \label{fig:donorkinetics}
\end{figure*}

\subsection{Fluorescence kinetics}
\label{fluorescence_kin}

In the above, we characterized the excitation energy transfer by a
single rate, which is possible if we assume fast equilibration
inside the donor and the acceptor band, and neglect the back
transfer from the acceptor to the donor. In general, one cannot
rely on these assumptions. As we mentioned in Sec.~\ref{strategy}
(also see Sec.~\ref{exact}), the most straightforward way to
characterize the energy transfer between donor and acceptor in
experiment is to follow the fluorescence kinetics of both
subsystems. In this section, we will use this approach. In all
cases, the calculations were done using the nonperturbative
method, i.e., accounting for band mixing and for relaxation within
and between the mixed bands. We will not assume a priori that
equilibration takes place on a timescale shorter than the one for
energy transfer. As only the bottom ($k=0$) states of the mixed
bands have oscillator strength, both the absorption and the
fluorescence spectrum consist of a single $\delta$-peak. Thus,
the total donor and acceptor fluorescence intensity as a function
of time are given by $I_D(t) = \gamma_{0+}P_{0+}(t)$ and $I_A(t) =
\gamma_{0-}P_{0-}(t)$, respectively. These are the quantities that
will be analyzed in the following and related to the intraband
relaxation and excatitation energy transfer.

\begin{figure*}[thb]
\centerline{\scalebox{0.95}{\includegraphics{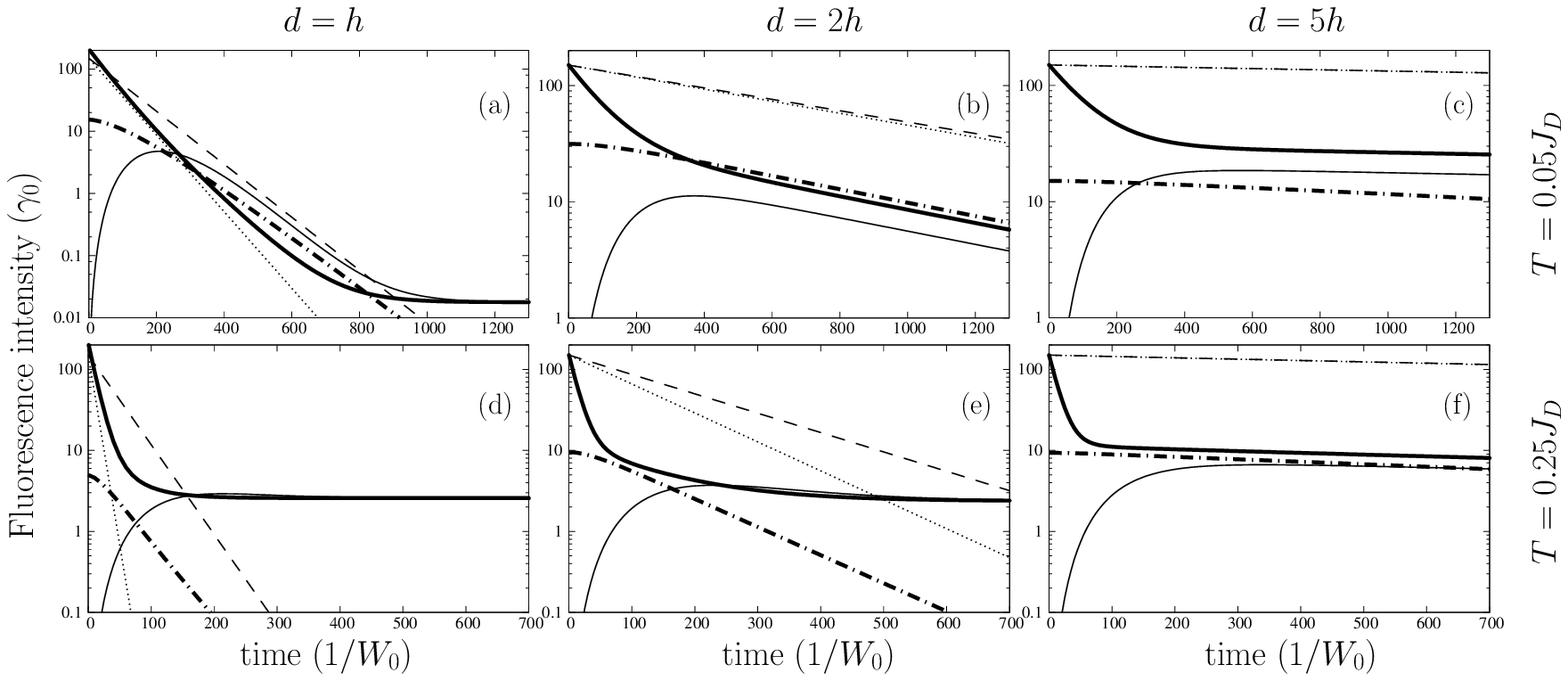}}}
\caption{Kinetics of the donor fluorescence for resonant
excitation (thick solid curve) and off-resonant excitation (thin
solid curve) calculated for various distances ($d/h = 1, 2$ and 5
from left to right) and temperatures ($T = 0.05 J_D$ and $0.25
J_D$ for top and bottom rows, respectively). All other parameters
as given at the beginning of Sec.~\ref{discussion}. The thick
dash-dotted curve shows the difference between the acceptor's
fluorescence intensity and its thermal equilibrium value (see
text). Finally, the straight lines represent the donor decay
assuming fast equilibration over the donor band in the
nonperturbative case [dashed curve, Eq.~(\ref{W+to-})] and the
perturbative limit [dotted curve, Eq.~(\ref{W+to-pert})].}
\label{fig:res-offres}
\end{figure*}

\subsubsection{Resonant excitation}
\label{resonant}

We first study the fluorescence kinetics for resonant excitation,
i.e., assuming that initially the superradiant state $|0+ \rangle$
of the donor band is excited by the pump. Following this
excitation, the population of this state has two channels to
relax, namely scattering to higher states in the donor band
(intraband relaxation) or scattering to the acceptor band (energy
transfer). Once transferred to the acceptor band, the excitation
may undergo relaxation within that band, or it may transfer back
to the donor. Figure~\ref{fig:donorkinetics} shows the donor
fluorescence kinetics $I_D(t)$ resulting from the interplay of
these processes at different temperatures and interchain
distances, calculated by solving Eq.~(\ref{Pnu}) with the above
initial condition. In doing so, we neglected the radiative decay
rates $\gamma_\nu$ of the exciton states, assuming this relaxation
channel to be much slower than all the others. As a result, at
nonzero temperatures, $P_{0+}(t)$ eventually reaches a finite
value, which is in accordance with the Boltzmann equilibrium over
donor and acceptor states. We notice that even the neglect of the
superradiant emission rate for chains of the order of $100$
molecules ($10-100\,$ps lifetime) does not limit the validity of
the results presented in this section. Typical values for $1/W_0$
are in the $1-10\,$fs time scale,~\cite{Bednarz03} so that at all
times considered in Figs.~\ref{fig:donorkinetics} and
\ref{fig:res-offres} the radiative decay indeed has a negligible
effect.

In Fig.~\ref{fig:donorkinetics} (b), (c), and (d) one can
distinguish two stages in the donor fluorescence kinetics, prior to
reaching the Boltzmann equilibrium. These stages are characterized
by different time scales. The first stage, indicated by I in
Fig.~\ref{fig:donorkinetics}(c), is a fast intraband relaxation.
Here, the initially created population of the superradiant donor
state $|0,+ \rangle$ rapidly scatters to the higher-lying dark
states $|k, + \rangle$, resulting in a fast decay of the donor's
fluorescence intensity. Obviously, this process gets more
pronounced with increasing temperature. The second stage,
indicated by II in Fig.~\ref{fig:donorkinetics}(c), is noticeably
slower. This stage is associated with the interband relaxation,
i.e., with the energy transfer between the donor and acceptor
bands. We note that at $T=0$ the upward intraband relaxation is
absent and the monoexponential fluorescence decay directly reflects
the transfer rate calculated through Eq.~(\ref{W+to-}) at $T=0$.

As is seen, stages I and II are easily separated as long as the
interchain distance $d > h$; it is then possible to determine a
meaningful effective energy transfer rate from the stage II decay
of the donor fluorescence. At $d = h$
[Fig.~\ref{fig:donorkinetics}(a)], it is hardly possible to
distinguish the two stages; the fluorescence kinetics only
exhibits a fast initial drop, almost directly followed by the
Boltzmann plateau. The reason is that for $d \lesssim h$, the bare
donor and acceptor bands are strongly mixed (see
Fig.~\ref{fig:pert-nonpert}). As a result, the intra- and
interband relaxation rates are of the same order, making it
impossible to distinguish their signatures in the fluorescence
kinetics. We also observe that the intraband relaxation is much
more sensitive to changing the temperature than the energy transfer
process. This is due to the fact that this relaxation is directly
sensitive to the number of thermally accessible dark states in the
donor band.

The above statement that the stage II part of the donor kinetics
indeed yields a meaningful measure for the energy transfer rate to
the acceptor, is corroborated by Fig.~\ref{fig:res-offres}, in
which for three different separations at two different
temperatures the donor fluorescence kinetics is plotted again
(thick solid curves), and is compared to the monoexponential decay
obtained through Eq.~(\ref{W+to-}) at the same temperature (dashed
line). In Figs.~\ref{fig:res-offres}(b,c,e,f) stage II can be
distinguished and this part of the kinetics is (roughly) parallel
to the monoexponential curve. The agreement deteriorates with
decreasing separation $d$ and with growing temperature. In panel
(e) stage II already is rather hard to see and the agreement with
the monoexponential curve is only fair. In panels (a) and (d),
stage II cannot be distinguished, and no part of the fluorescence
kinetics shows a reasonable agreement with the monoexponential
decay line.

From the above it may be concluded that in the case of strong
interchain coupling, under conditions of resonant excitation, the
kinetics of the donor fluorescence is not a good tool to determine
the energy transfer rate. One may expect that under these
conditions, the acceptor fluorescence yields a better tool,
because this quantity is influenced less directly by the upward
relaxation within the donor band. For this reason, we also plotted
in Fig.~\ref{fig:res-offres} the acceptor fluorescence as a
function of time (thick dash-dotted curves). These curves represent
$\gamma_{0-}[P_{0-}(t \to \infty) - P_{0-}(t)]$, where $P_{0-}(t
\to \infty)$ equals the thermal equilibrium value. We observe that
even in the case of strong coupling and high temperature, the
acceptor fluorescence indeed seems to be affected less by the
initial thermal relaxation part and more directly reflects the
energy transfer rate as calculated assuming equilibration (dashed
lines).

\subsubsection{Off-resonant excitation}
\label{off-resonant}

We next turn to the case of off-resonant excitation. To this end
we also plotted in Fig.~\ref{fig:res-offres} (thin solid curve) the
donor fluorescence after the system has been brought
(artificially) in the initial state with $P_{k+}(t)=\delta_{k,N/2}$
and $P_{k-}(t)=0$, i.e., the center state of the donor band is
excited. As for the case of resonant excitation, we observe that
at low temperatures and (or) large interchain separations, the
kinetics exhibits two stages prior to reaching the Boltzmann
plateau. The first stage, intraband relaxation, now manifests
itself through the increase of the donor fluorescence from its
initial value zero, because the donor population needs time to
reach the superradiant bottom state of the donor band. We note
that, in spite of the many relaxation steps needed to reach the
bottom, for low temperatures this first stage occurs on the same
timescale or even faster than for the case of resonant excitation.
As a result the stage II process, related to the energy transfer,
reflects the same timescale as for resonant excitation.

By contrast, at higher temperatures the distinction between the
first and second stage of the kinetics is less clear: there is a
more gradual change between the two, which also creates the
impression of a longer timescale for the intraband relaxation. We
explain this as follows. As near the band center the mixing of
donor and acceptor bands is very strong, the intraband and
interband relaxation there happen at the same timescale. As a
result, off-resonant excitation is followed by very fast
distribution of the excitation over donor and acceptor bands, in
both of which the population relaxes to the bottom. Thus, as
opposed to the case of resonant excitation, the energy transfer
problem at the bottom of the band takes place in a situation where
a large part of the population already has ended up in the
acceptor band. In this situation, the back-transfer from the
acceptor to the donor, which takes place at temperatures of the
order of the band edge detuning $\Delta$, feeds the donor
fluorescence after the initial relaxation, which prolongs the time
during which the donor fluorescence increases and slows down its
overall kinetics. Comparing to the dashed lines in
Fig.~\ref{fig:res-offres}, we observe that at the higher
temperatures no part of the kinetics curve for off-resonant
excitation reflects the energy transfer rate calculated through
Eq.~(\ref{W+to-}). Thus, at high temperatures (comparable to
$\Delta$ or higher) off-resonantly excited donor fluorescence seems
not to be a good ruler for the energy transfer rate. At low
temperatures, this technique is more useful and then yields the
same transfer rate as donor or acceptor
fluorescence after resonant excitation of the donor.

Without showing details, we mention that in the case of
off-resonant excitation also the kinetics of the acceptor
fluorescence suffers from the interplay between intraband and
interband transfer and therefore also does not yield a proper tool
for measuring the energy transfer rate.

\section{Summary and conclusions}
\label{summary}

We theoretically studied the excitation energy transfer between two
linear chains of dye molecules, whose primary excitations are
Frenkel excitons. The superradiant band bottoms of both chains were
detuned by a value $\Delta$; the chain with the higher (lower)
band bottom was considered the energy donor (acceptor). In addition
to electronic interchain interactions, our theory accounts for weak
exciton-phonon coupling. Two methods were investigated. In the
perturbative method, we used the common approximation of treating
the interchain interactions as a perturbation, which also is the
basis for the standard FRET formalism. In the nonperturbative
approach, we allowed for mixing of the donor and acceptor exciton
bands and considered the various phonon-induced relaxation
mechanisms in and between these bands to describe the energy
transfer process.

Using the perturbative approach, we have confirmed that energy
transfer between multichromophoric systems can occur from and
towards dipole forbidden states.\cite{Sumi99,%
Mukai99,Scholes01,Jordanidis01,Scholes03,Jang04,Fleming04,Fukutake02}
We have focused, however, on two other aspects specific to
multichromophoric systems, namely the effects of band mixing and
intraband relaxation on the energy transfer. The first has been
studied by comparison of the perturbative and the nonperterbative
method. Not surprisingly, we find that the perturbative approach
looses its validity when the interchain distance decreases (and
the interchain interactions grow). The perturbative result for the
energy transfer rate then strongly overestimates the exact one
that accounts for band mixing. The distance at which the
perturbation theory breaks down increases with temperature,
because the effect of band mixing is larger for the states inside
the bands, which get populated upon heating. We also compared our
results to standard FRET theory, which turns out to lose its
validity as soon as the interchain separation falls below the
chain length (or exciton delocalization length, in case one
considers disordered chains).

From our study it turns out that the process of energy transfer
between multichromophoric systems is strongly affected by
intraband relaxation. This is most obvious in the kinetics of the
donor and acceptor fluorescence after pulsed excitation. As these
quantities are natural choices for probing energy transfer, it is
important to realize that this kinetics does not only reflect the
actual transfer process of interest, but also is affected by
thermal relaxation inside the donor and acceptor bands. Only if
both processes can be separated, due to different time scales, one
may extract the energy transfer rate from such experiments. Our
results suggest that the best way of determining the transfer rate
between two J-aggregates is to measure the fluorescence kinetics
of the acceptor's J-band after resonant excitation of the donor
band. Under these conditions, the approach of the acceptor's
fluorescence intensity grows towards its equilibrium value almost
mono-exponentially, allowing for a meaningful definition of the
transfer rate. We also found that the thus extracted rate agrees
well with the transfer rate obtained when assuming equilibration
over the donor band. If, due to relaxation to the ground state,
measuring the acceptor's equilibrium fluorescence intensity turns
out to be a problem, one may also resort to analyzing the decay of
the time-derivative of this intensity. The first component of that
decay should then reflect the transfer rate, while the second
reflects the relaxation to the ground state.

We note that throughout this paper we have neglected the homogeneous
linewidths $\Gamma_k$ of the individual exciton transitions. Within
our model, these widths result from exciton-phonon scattering.~\cite{Heijs05}
It may be estimated that within the context of excitation energy
transfer these widths may indeed be neglected (i.e., $\Gamma_k \ll J_{kk}^{DA}$)
as long as $W_0 \ll 2\pi J_D^2/\omega_c$.

The results of this paper were derived using the simplest possible
model of two interacting homogeneous chains with periodic boundary
conditions. The same issues of band mixing and intraband
relaxation will play an important role for more general systems,
where one allows for energy and interaction disorder and (or) for
different geometries. In fact, all expressions in
Secs.~\ref{perturbative} and \ref{exact} also hold for such more
general situations, provided that the various wave functions
$\varphi_{kn}^D$, $\varphi_{kn}^A$, or $\phi_{\mu n}$ as well as
the energies $E_k^D$, $E_k^A$, or $E_\mu$ are replaced by the
eigenvectors and eigenenergies of the corresponding exciton
Hamiltonian. An example of great interest to analyze next, is a
pair of concentric cylindrical aggregates. As explained in the
Introduction, our model of two parallel linear aggregates was
inspired on this experimentally realized
situation.~\cite{Didraga04,Pugzlys04} Varying the distance between
both walls in this system is possible by altering molecular side
groups and (or) changing the solvent.

\acknowledgments This research is supported by NanoNed, a national
nanotechnology programme coordinated by the Dutch Ministry of
Economic Affairs.

\appendix
\section{Some explicit expressions for homogeneous chains}

In this Appendix, we present analytical
expressions for the probability overlap functions valid for the
ordered case.

Substituting the eigenfunctions Eq.~(\ref{eigenstates}) into
Eq.~(\ref{overlap}), the probability overlap ${\mathcal O}_{k
k^{\prime}}^X$ is obtained as
\begin{equation}
    {\mathcal O}_{k k^{\prime}}^X = \frac{1}{N} \ .
\end{equation}
Likewise, from Eq.~(\ref{pm}) and Eq.~(\ref{overlapD+A}) we obtain
the probability overlap ${\mathcal O}_{\mu\nu}$ [with $\mu$ and
$\nu$ chosen from $(k,+)$ or $(k,-)$] in the form
\begin{equation}
\label{Omunu}
    {\mathcal O}_{\mu\nu}= \frac{1}{2N} \begin{cases}
    1+\dfrac{1}{\sqrt{(1+\eta_k)(1+\eta_{k^{\prime}})}} \ ,
    & \mu=(k,\pm) \ ,
    \\
    & \nu=(k^{\prime},\pm)\ ,
    \\
    \\
    1-\dfrac{1}{\sqrt{(1+\eta_k)(1+\eta_{k^{\prime}})}} \ ,
    & \mu=(k,\pm) \ ,
    \\
    & \nu=(k^{\prime},\mp) \ .
\end{cases}
\end{equation}


\begin{thebibliography}{99}


\bibitem{Foerster48} Th.\ F\"orster, Ann. Phys.\ \textbf{2}, 55
    (1948); in {\it Modern Quantum Chemistry}, Pt. III, edited by
    O. Sinanoglu (Academic, New York, 1965).

\bibitem{Dexter53} D.\ L.\ Dexter, J.\ Chem.\ Phys.\ {\bf 21},
    836 (1953).

\bibitem{Ermolaev77} V.\ L.\ Ermolaev, E. N. Bodunov, E. B.
    Sveshnikova, and T. A. Shakhverdov, {\it Nonradiative Energy
    Transfer of Electronic Excitation} (Nauka, Leningrad, 1977),
    in Russian.

\bibitem{Agranovich82} V.\ M.\ Agranovich and M. D. Galanin,
    {\it Electronic Excitation Energy Transfer in Condensed
    Matter} (North-Holland, Amsterdam, 1982).

\bibitem{Andrews99}{\it Resonance Energy Transfer}, edited by
    D. L. Andrews and A. A. Demidov (Wiley, Chichester, 1999).

\bibitem{Ermolaev96} V. L. Ermolaev, E. B. Sveshnikova, and E. N.
    Bodunov, Usp Fiz. Nauk {\bf 166}, 279 (1996)[Phys. Uspekhi
    {\bf 39}, 261 (1996)].

\bibitem{Jang02} S. Jang, Y. J. Jung, R. J. Silbey, Chem.
    Phys. {\bf 275}, 319 (2002).

\bibitem{Kagan96} C. R. Kagan, C. B. Murray, and M. G. Bawendy,
    Phys. Rev. B {\bf 54}, 8633 (1996).

\bibitem{Crooker02} S. A. Crooker, J. A. Hollingdworth, S. Tretiak,
    and V. L. Klimov, Phys. Rev. Lett. {\bf 89}, 186802 (2002).

\bibitem{Wargnier04} R. Wargnier, A. V. Baranov, V. G. Maslov,
    V. Stsiapura, M. Artemyev, M. Pluot, A. Sukhanova, and I.
    Nabiev, Nano Lett. {\bf 4}, 451 (2004).

\bibitem{Javier03} A. Javier, C. Steven Yun, J. Sorena, and G. F.
    Strose, J. Phys. Chem. B {\bf 107} (2003).

\bibitem{Imamoglu99} A. Imamo\~glu, D. D. Awschalom, G. Burkard,
    D. P. DiVincenzo, D. Loss, M. Sherwin, and A. Small, Phys.
    Rev. Lett, {\bf 83}, 4204 (1999).

\bibitem{Biolatti00} E. Biolatti, R. C. Iotti, P. Zanardi, and
    F. Rossi, Phys. Rev. Lett. {\bf 85}, 5647 (2000).

\bibitem{Holstein81}T. Holstein, S. K. Lyo, and R. Orbach, in
    {\it Laser Spectroscopy of Solids}, edited by W. M. Yen and
    P. M. Selzer (Springer, Berlin - Heidelberg - New York, 1981),
    p. 39.

\bibitem{Malyshev85}V. A. Malyshev, in {\it Spectroscopy of
    Crystalls}, edited by A. A. Kaplyanskii (Nauka, Leningrad,
    1985), p. 100, in Russian.

\bibitem{Basiev87} T.\ T.\ Basiev, V. A. Malyshev, and A. K.
    Przhevuskii, in {\it Spectroscopy of Solids Containing
    Rare-Earth Ions}, edited by A. A. Kaplyanskii and R. M.
    Macfarlane (North-Holland, Amsterdam, 1987), p. 275.

\bibitem{Xia02}S. Xia and P. A. Tanner, Phys. Rev. B {\bf 66},
    214305 (2002).

\bibitem{Sumi99} H. Sumi, J. Phys. Chem. B {\bf 103}, 252 (1999).

\bibitem{Mukai99} K. Mukai, S. Abe, and H. Sumi, J. Phys. Chem. B
    {\bf 103}, 6096 (1999).

\bibitem{Scholes01} G. D. Scholes, X. J. Jordanidis, and G. R.
    Fleming, J. Phys. Chem. B {\bf 105}, 1640 (2001).

\bibitem{Jordanidis01} X. J. Jordanidis, G. D. Scholes, and G. R.
    Flemming, J. Phys. Chem. B {\bf 105}, 1640 (2001).

\bibitem{Scholes03} G. D. Scholes, Annu. Rev. Phys. Chem.
    {\bf 54}, 57 (2003).

\bibitem{Jang04} S. Jang, M. D. Newton, R. J. Silbey, Phys. Rev.
    Lett. {\bf 92}, 218301 (2004).

\bibitem{Fleming04} G. R. Fleming, G. D. Scholes, Nature {\bf 431},
    256 (2004).

\bibitem{Shreve91} A. P. Shreve, J. K. Trautman, H. A. Frank, T.G.
    Owens, and A. C. Albrecht, Biochim. Biophys. Acta {\bf 1058},
    280 (1991).

\bibitem{Jimenez96} R. Jimenez, S. N. Dikshit, S. E. Bradforth,
    G. R. Fleming, J. Phys. Chem. {\bf 100}, 6825 (1996).

\bibitem{vanOijen99} A. M. van Oijen, M. Ketelaars, J. K\"oler,
    T. J. Aartsma, and J. Schmidt, Science {\bf 285}, 400 (1999).

\bibitem{Wu97} H.-M. Wu, M. Ratsep, R. Jankowiak, R. J. Cogdell,
    and G. J. Small, J. Phys. Chem. B {\bf 101}, 7641 (1997).

\bibitem{vonBerlepsch00a} H.\ von Berlepsch, C.\ B\"ottcher, A. Quart,
    M.\ Regenbrecht, S.\ Akari, U.\ Keiderling, H.\ Schnablegger,
    S. D\"ahne, and S.\ Kirstein, Langmuir {\bf 16}, 5908 (2000).

\bibitem{vonBerlepsch00b} H.\ von Berlepsch, C.\ B\"ottcher, A. Quart,
    C.\ Burger, S. D\"ahne, and S.\ Kirstein, J. Phys. Chem. B {\bf 104},
    5255 (2000).

\bibitem{vonBerlepsch03} H.\ von Berlepsch, S.\ Kirstein, R.\ Hania,
    C.\ Didraga, A. Pu\v{z}glys, and C.\ B\"ottcher, J. Phys. Chem. B
    {\bf 107}, 14176 (2003).

\bibitem{Didraga04} C. Didraga, A. Pug\v{z}lys, P. R. Hania, H.
    von Berlepsch, K. Duppen, and J. Knoester, J. Phys. Chem. B
    {\bf 108}, 14976 (2004).

\bibitem{Pugzlys04} A. Pug\v{z}lys, P. R. Hania, C. Didraga, V. A.
    Malyshev, J. Knoester, and K. Duppen, in {\it Ultafast Phenonena
    XIV}, edited by T. Kobayashi, T. Okada, T. Kobayashi, K. A. Nelson,
    and S. De Silvestri (Springer, Berlin - Heidelberg - New York,
    2005), Springer Series in Chemical Physics, vol. 79, p. 879.

\bibitem{Fukutake02} N. Fukutake, S. Takasaka, and T. Kobayashi,
    Chem. Phys. Lett. {\bf 361}, 42 (2002).

\bibitem{Heijs05} D. J. Heijs, V. A. Malyshev, and J. Knoester,
    Phys. Rev. Lett. {\bf 95}, 177402 (2005);
    J. Chem. Phys. {\bf 123}, 144507 (2005).

\bibitem{Bednarz02} M. Bednarz, V.A. Malyshev, J. Knoester, J. Chem.
    Phys. {\bf 117}, 6200 (2002); {\bf 120}, 3827 (2004).

\bibitem{LandauQM} L. D. Landau and Lifshits, {\it Quantum Mechanics}
    (Pergamon, Oxford, 1965).

\bibitem{Weiss}  U. Weiss,  {\it Quantum Dissipative Systems}
     (World Scientific, Singapore, 1993).

\bibitem{Kuhn97} O. K\"uhn and V. Sundstr\"om, J. Chem. Phys. B {\bf 107},
    4154 (1997).

\bibitem{May00} V. May and O. K\"uhn, {\it Charge and Energy Transfer
    Dynamics in Molecular Systems} (Wiley-VCH, Berlin, 2000).

\bibitem{Renger01} T. Renger, V. May, and O. K\"uhn, Phys. Rep. {\bf
    343}, 137 (2001).

\bibitem{Brueggemann04} B. Br\"uggemann, K. Szenee, V. Novoderzhkin,
    R. van Grondelle, and V. May, J. Phys. Chem. B {\bf 108},
    13536 (2004).

\bibitem{Bednarz03} M. Bednarz, V.A. Malyshev, J. Knoester, Phys.
    Rev. Lett. {\bf 91}, 217401 (2003).


\end{thebibliography}
\end{document}